# Negative Observations in
# Quantum Mechanics

Douglas M. Snyder

Quantum mechanics is fundamentally a theory concerned with knowledge of the physical world.  It is not fundamentally concerned with describing the functioning of the physical world independent of the observing, thinking person, as Newtonian mechanics is generally considered to be (Snyder, 1990, 1992).  Chief among the reasons for the thesis that cognition and the physical world are linked in quantum mechanics is that all knowledge concerning physical existents is developed using their associated wave functions, and the wave functions provide only probabilistic knowledge regarding the physical world (Liboff, 1993).  There is no physical world in quantum mechanics that is assumed to function independently of the observer who uses quantum mechanics to develop predictions and who makes observations that have consistently been found to support these predictions.  Also significant is the immediate change in the quantum mechanical wave function associated with a physical existent that generally occurs throughout space upon measurement of the physical existent.  This change in the wave function is not limited by the velocity limitation of the special theory of relativity for physical existents–the velocity of light in vacuum.

Another relevant feature of quantum mechanics is the complex number nature of the wave function associated with a physical existent that is the basis for deriving whatever information can be known concerning the existent (Eisberg & Resnick, 1974/1985).  A complex function is one that has both mathematically imaginary and real components.  The physical world is traditionally described by mathematically real numbers, giving rise to Eisberg and Resnick's (1974/1985) comment that "we should not attempt to give to wave functions [in quantum mechanics] a physical existence in the same sense that water waves have a physical existence" (p. 147).

Nonetheless, the particular demonstration concerning the phenomenon of interference to be discussed in the next section is remarkable.  Examining interference will spotlight the wave-particle duality in quantum mechanics, the key feature of this duality being that physical existents sometimes show particle-like characteristics and sometimes show wave-like characteristics.  Wave functions exhibiting interference are based on the sum of two or more elementary wave functions.  In contrast, where interference does not characterize some physical phenomenon, this phenomenon is described by a wave function that consists of only one of these elementary wave functions.

## Feynman's Two-Hole Gedankenexperiments

Generally the change in the wave function that often occurs in measurement in quantum mechanics has been ascribed to the unavoidable physical interaction between the measuring instrument and the physical entity measured. Indeed, Bohr (1935) maintained that this unavoidable interaction was responsible for the uncertainty principle, more specifically the inability to simultaneously measure observable quantities described by non-commuting Hermitian operators (e.g., the position and momentum of a particle). The following series of gedankenexperiments in this section will show that this interaction is not necessary to effect a change in the wave function. The series of gedankenexperiments indicates that knowledge plays a significant role in the change in the wave function that often occurs in measurement (Snyder, 1996a, 1996b).

### *Gedankenexperiment 1*

Feynman, Leighton, and Sands (1965) explained that the distribution of electrons passing through a wall with two suitably arranged holes to a backstop where the positions of the electrons are detected exhibits interference (Figure 1). Electrons at the backstop may be detected with a Geiger counter or an electron multiplier. Feynman et al. explained that this interference is characteristic of wave phenomena and that the distribution of electrons at the backstop indicates that each of the electrons acts like a wave as it passes through the wall with two holes. It should be noted that when the electrons are detected in this gedankenexperiment, they are detected as discrete entities, a characteristic of particles, or in Feynman et al.'s terminology, "lumps" (p. 1-5).

In Figure 1, the absence of lines indicating possible paths for the electrons to take from the electron source to the backstop is not an oversight. An electron is not taking one or the other of the paths. Instead, the wave function associated with each electron after it passes through the holes is the sum of two more elementary wave functions, with each of these wave functions experiencing diffraction at one or the other of the holes. Epstein (1945) emphasized that when the quantum mechanical wave of some physical entity such as an electron exhibits interference, it is interference generated only in the wave function characterizing the individual entity.

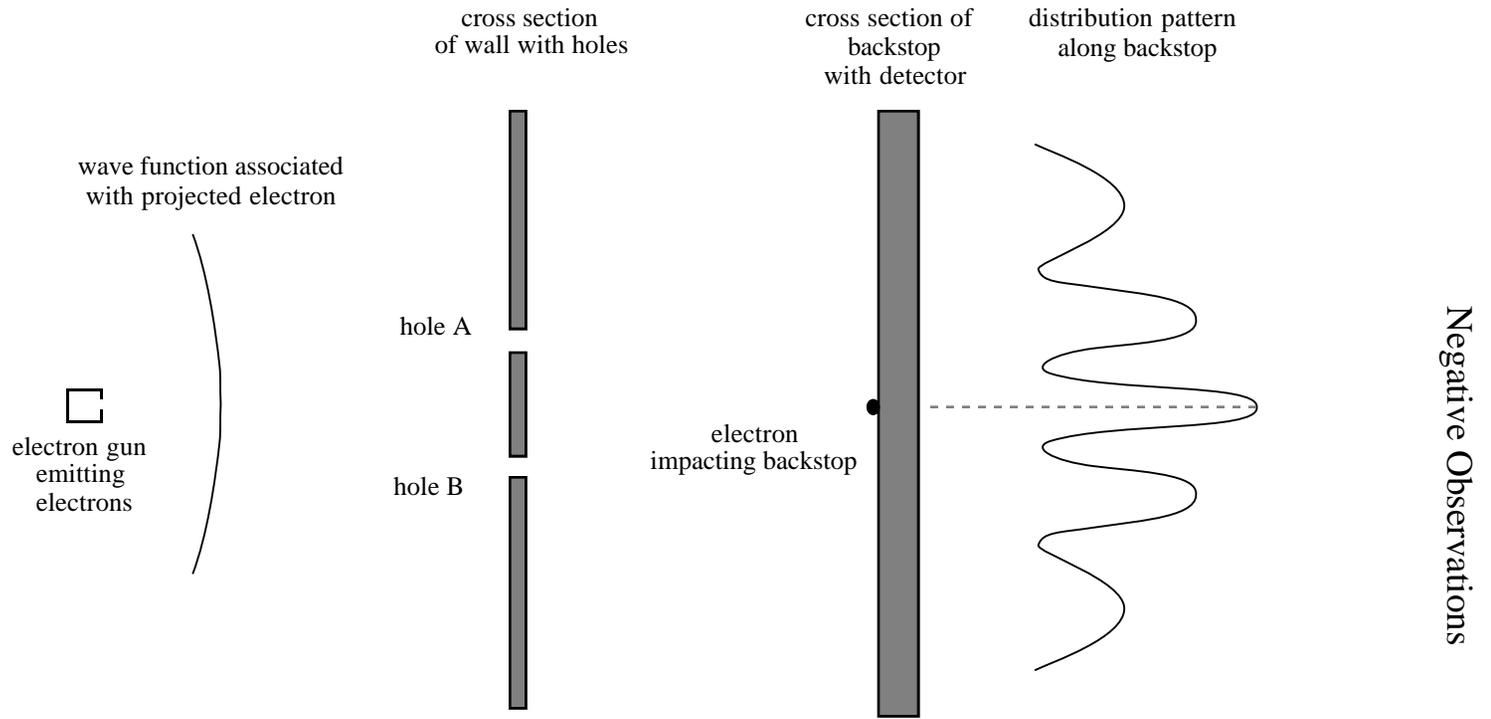

cross section
of wall with holes

cross section of
backstop
with detector

distribution pattern
along backstop

wave function associated
with projected electron

hole A

hole B

electron gun
emitting
electrons

electron
impacting backstop

Negative Observations

Figure 1

Two-hole gedankenexperiment in which the distribution of
electrons reflects interference in the wave functions of electrons.
(Gedankenexperiment 1)

# Negative Observations

The diffraction patterns resulting from the waves of the electrons passing through the two holes would at different spatial points along a backstop behind the hole exhibit constructive or destructive interference. At some points along the backstop, the waves from each hole sum (i.e., constructively interfere), and at other points along the backstop, the waves from each hole subtract (i.e., destructively interfere). The distribution of electrons at the backstop is given by the absolute square of the combined waves at different locations along the backstop, similar to the characteristic of a classical wave whose intensity at a particular location is proportional to the square of its amplitude. Because the electrons are detected as discrete entities, like particles, at the backstop, it takes many electrons to determine the intensity of the quantum wave that describes each of the electrons and that is reflected in the distribution of the electrons against the backstop.

*Gedankenexperiment 2*

Feynman et al. further explained that if one were to implement a procedure in which it could be determined through which hole the electron passed, the interference pattern is destroyed and the resulting distribution of the electrons resembles that of classical particles passing through the two holes in an important way. Feynman et al. relied on a strong light source behind the wall and between the two holes that illuminates an electron as it travels through either hole (Figure 2). Note the significant difference between the distribution patterns in Figures 1 and 2.

In Figure 2, the path from the electron's detection by the light to the backstop is indicated, but it is important to emphasize that this path is inferred only after the electron has reached the backstop. A measurement of the position of the electron with the use of the light source introduces an uncertainty in its momentum. Only when the electron is detected at the backstop can one infer the path the electron traveled from the hole it went through to the backstop. It is not something one can know before the electron strikes the backstop.

In Feynman et al.'s gedankenexperiment using the light source, the distribution of electrons passing through both holes would be similar to that found if classical particles were sent through an analogous experimental arrangement in an important way. Specifically, as in the case of classical particles, this distribution of electrons at the backstop is the simple summation of the distribution patterns for electrons passing through one or the other of the holes. Figure 3 shows the distribution patterns of electrons passing through

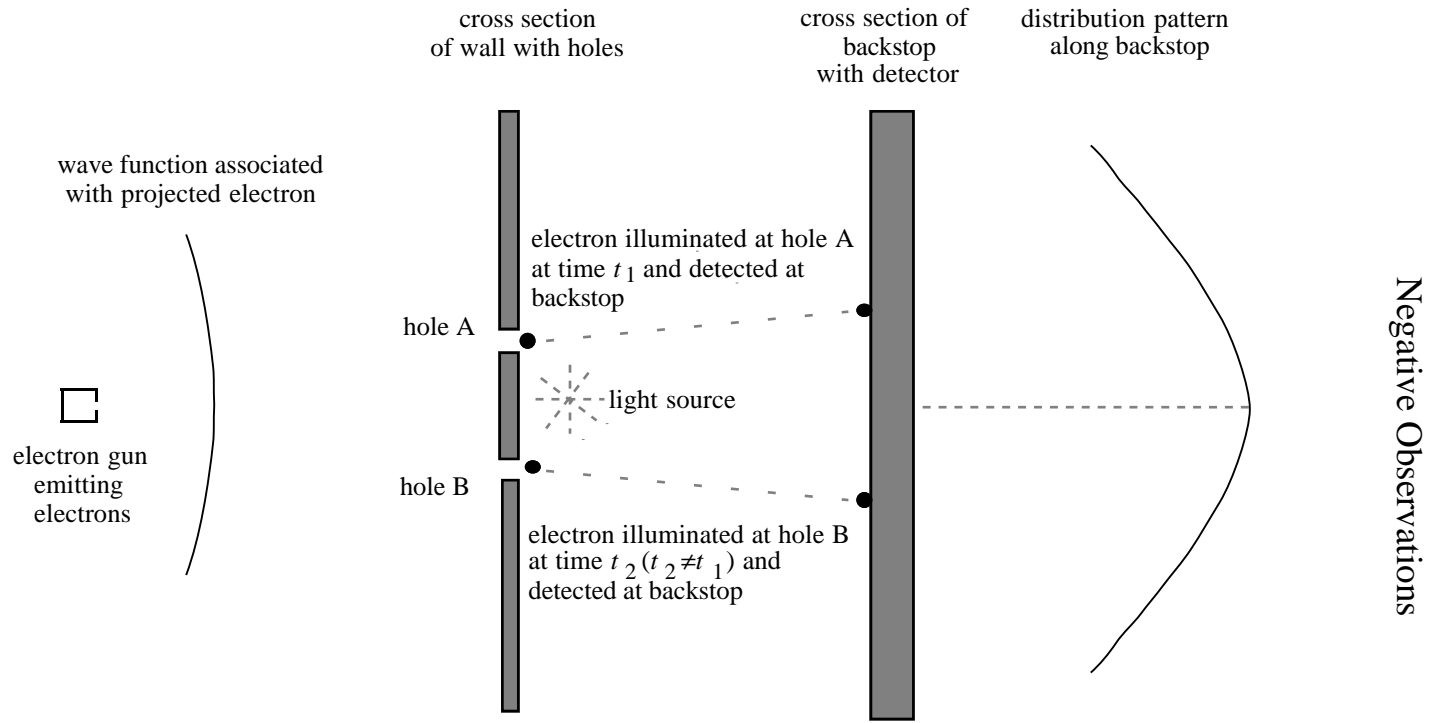

Figure 2
Two-hole gedankenexperiment with strong light source.
(Gedankenexperiment 2)

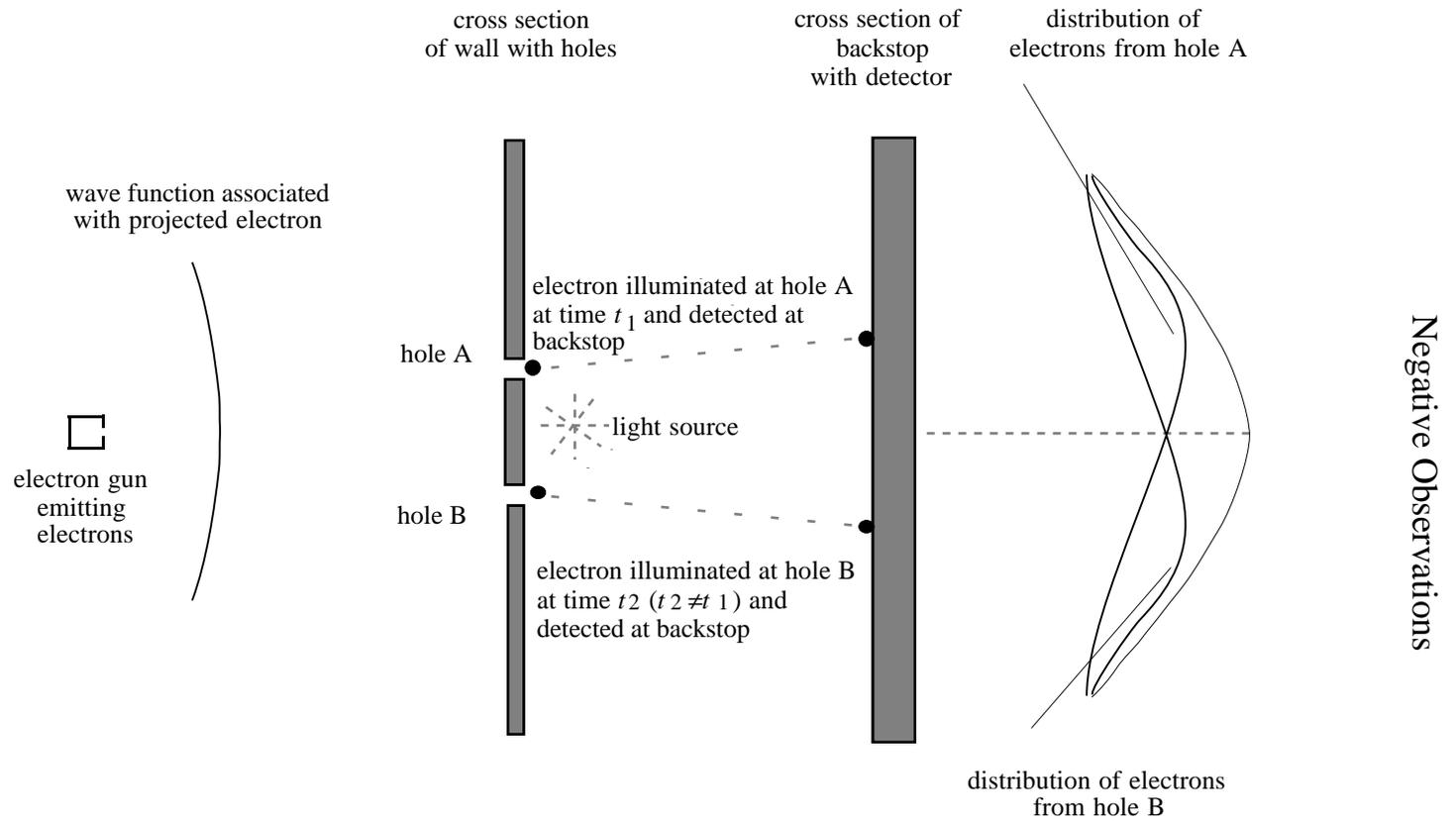

Figure 3

Two-hole gedankenexperiment with strong light source in which the
distribution of electrons from each hole is shown.

hole A and electrons passing through hole B in Gedankenexperiment 2. These distribution patterns are identical to those that would occur if only one or the other of the holes were open at a particular time. An inspection of Figure 3 shows that summing the distribution patterns for the electrons passing through hole A and those passing through hole B results in the overall distribution of electrons found in Gedankenexperiment 2.

*The Uncertainty Principle*

Feynman et al.'s gedankenexperiments are themselves very interesting in that they illustrate certain apparently incongruent characteristics of microscopic physical existents, namely particle-like and wave-like features. Feynman et al. discussed their gedankenexperiments in terms of Heisenberg's uncertainty principle. Feynman et al. wrote:

> He [Heisenberg] proposed as a general principle, his *uncertainty principle*, which we can state in terms of our experiment as follows: "It is impossible to design an apparatus to determine which hole the electron passes through, that will not at the same time disturb the electrons enough to destroy the interference pattern." If an apparatus is capable of determining which hole the electron goes through, it *cannot* be so delicate that it does not disturb the pattern in an essential way. (p. 1-9)

Note that Feynman et al. implied in their description of the uncertainty principle that there is an unavoidable interaction between the measuring instrument (in their gedankenexperiment, the strong light source emitting photons) and the physical entity measured. Feynman et al. also wrote concerning Gedanken-experiment 2:

> the jolt given to the electron when the photon is scattered by it is such as to change the electron's motion enough so that if it might have gone to where $P_{12}$ [the electron distribution] was at a maximum [in Gedankenexperiment 1] it will instead land where $P_{12}$ was at a minimum; that is why we no longer see the wavy interference effects. (p. 1-8)

In determining through which hole an electron passes, Feynman et al., like most physicists, maintained that the electrons are unavoidably disturbed by the photons from the light source and it is this disturbance by the photons that destroys the interference pattern. Indeed, in a survey of a number of the textbooks of quantum mechanics, it is interesting that each author, in line with Feynman and Bohr, allowed a central role in the change in the wave function that occurs in a measurement to a physical interaction between the physical existent measured and some physical measuring apparatus. The authors of these textbooks are Dicke and Witke (1960), Eisberg and Resnick (1974/1985), Gasiorowicz (1974), Goswami (1992), Liboff (1993), Merzbacher (1961/1970), and Messiah (1962/1965).

It is important to note explicitly that some causative factor is necessary to account for the very different distributions of the electrons in Figures 1 and 2. Feynman et al. maintained that the physical interaction between the electrons and photons from the light source is this factor.

*Gedankenexperiment 3*

Feynman et al.'s gedankenexperiments indicate that in quantum mechanics the act of taking a measurement in principle is linked to, and often affects, the physical world which is being measured. The nature of taking a measurement in quantum mechanics can be explored further by considering a certain variation of Feynman et al.'s second gedankenexperiment (Epstein, 1945; Renninger, 1960).[8] The results of this exploration are even more surprising than those presented by Feynman et al. in their gedanken-experiments. Empirical work on electron shelving that supports the next gedankenexperiment has been conducted by Nagourney, Sandberg, and Dehmelt (1986), Bergquist, Hulet, Itano, and Wineland (1986), and by Sauter, Neuhauser, Blatt, and Toschek (1986). This work has been summarized by Cook (1990).[9]

---

[8] Epstein (1945) presented the essence of Gedankenexperiment 3 using the passage of photons through an interferometer. Renninger (1960) also discussed a gedankenexperiment in an article entitled "Observations without Disturbing the Object" in which the essence of Gedankenexperiment 3 is presented.

[9] In electron shelving, an ion is placed into a superposition of two quantum states. In each of these states, an electron of the ion is in one or the other of two energy levels. The transition to one of the quantum states occurs very quickly and the transition to the other state occurs very slowly. If the ion is repeatedly placed in the superposition of states after it transitions to one or the other of the superposed states, one finds the atomic electron in general transitions very frequently between the superposed quantum states and the quantum state characterized by the very quick transition. The photons emitted in these frequently occurring transitions to the quantum state characterized by the very quick transition are associated with resonance fluorescence of the ion. The absence of resonance fluorescence means that the ion has transitioned into the quantum state that occurs infrequently.

Cook (1990) has pointed out that in the work of Dehmelt and his colleagues on electron shelving involving the Ba$^+$ ion, the resonance fluorescence of a single ion is of sufficient

# Negative Observations

In a similar arrangement to that found in Gedankenexperiment 2, one can determine which of the two holes an electron went through on its way to the backstop by using a light that is placed near only one of the holes and which illuminates only the hole it is placed by (Figure 4). Illuminating only one of the holes yields a distribution of the electrons similar to that which one would expect if the light were placed between the holes, as in Feynman et al.'s second gedankenexperiment. The distribution is similar to the sum of the distributions of electrons that one would expect if only one or the other of the holes were open at a particular time.

Moreover, when an observer knows that electrons have passed through the unilluminated hole because they were not seen to pass through the illuminated hole, the distribution of these electrons through the unilluminated hole resembles the distribution of electrons passing through the illuminated hole (Figure 5). Consider also the point that if: 1) the light is turned off before sufficient time has passed allowing the observer to conclude that an electron could not have passed through the illuminated hole, and 2) an electron has not been observed at the illuminated hole, the distribution of many such electrons passing through the wall is determined by an interference pattern that is the sum of diffraction patterns of the waves of the electrons passing through the two holes similar to that found in Gedankenexperiment 1 (Epstein, 1945; Renninger, 1960).

*Discussion of the Gedankenexperiments*

The immediate question is how are the results in Gedankenexperiment 3 possible given Feynman et al.'s thesis that physical interaction between the light source and electron is necessary to destroy the interference? Where the light illuminates only hole A, electrons passing through hole B do not interact with photons from the light source and yet interference is destroyed in the same manner as if the light source illuminated both holes A and B. In addition, the distribution of electrons passing through hole B at the backstop indicates that there has been a change in the description of these electrons, even though no physical interaction has occurred between these electrons and photons from the light source.

---

intensity to be detectable by the dark-adapted eye alone, and the making of a negative observation, to be discussed shortly, is thus not dependent on any measuring device external to the observer.

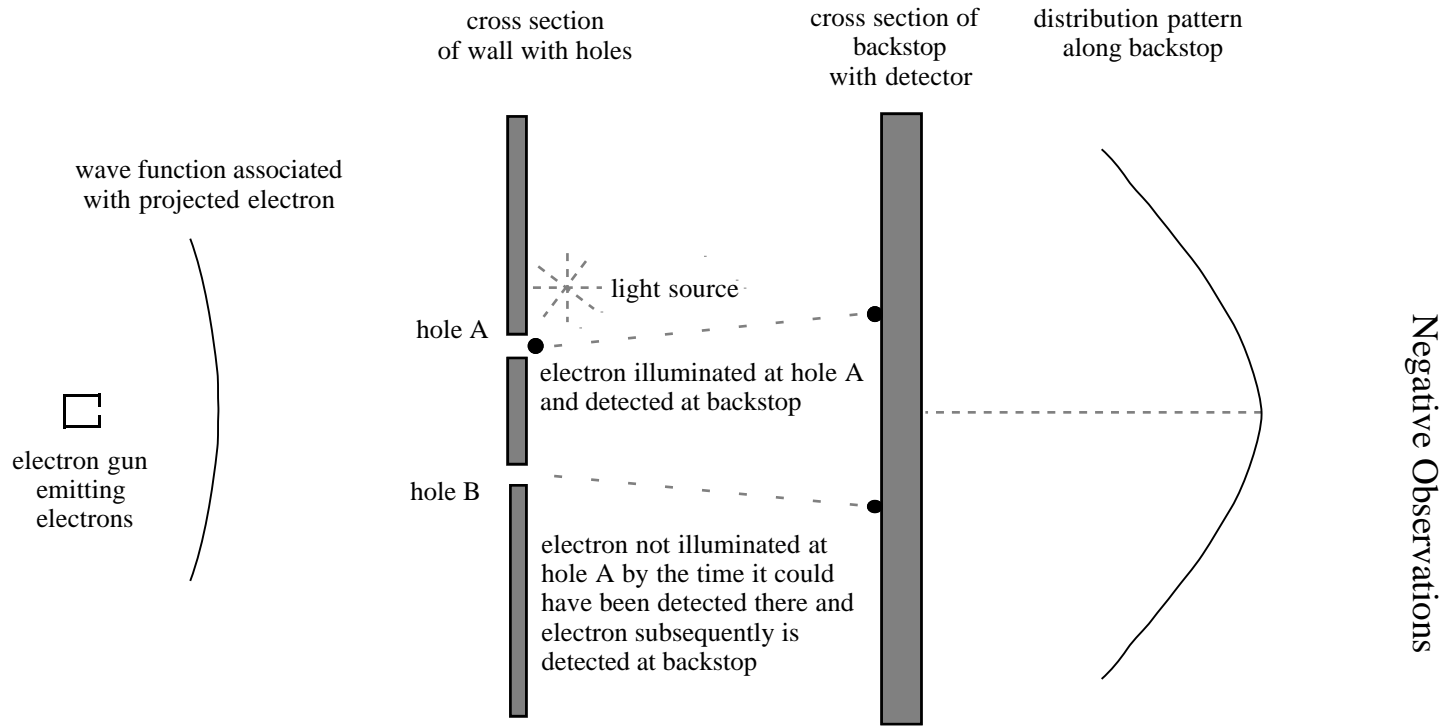

Figure 4

Two-hole gedankenexperiment with strong light source illuminating only one hole.
(Gedankenexperiment 3)

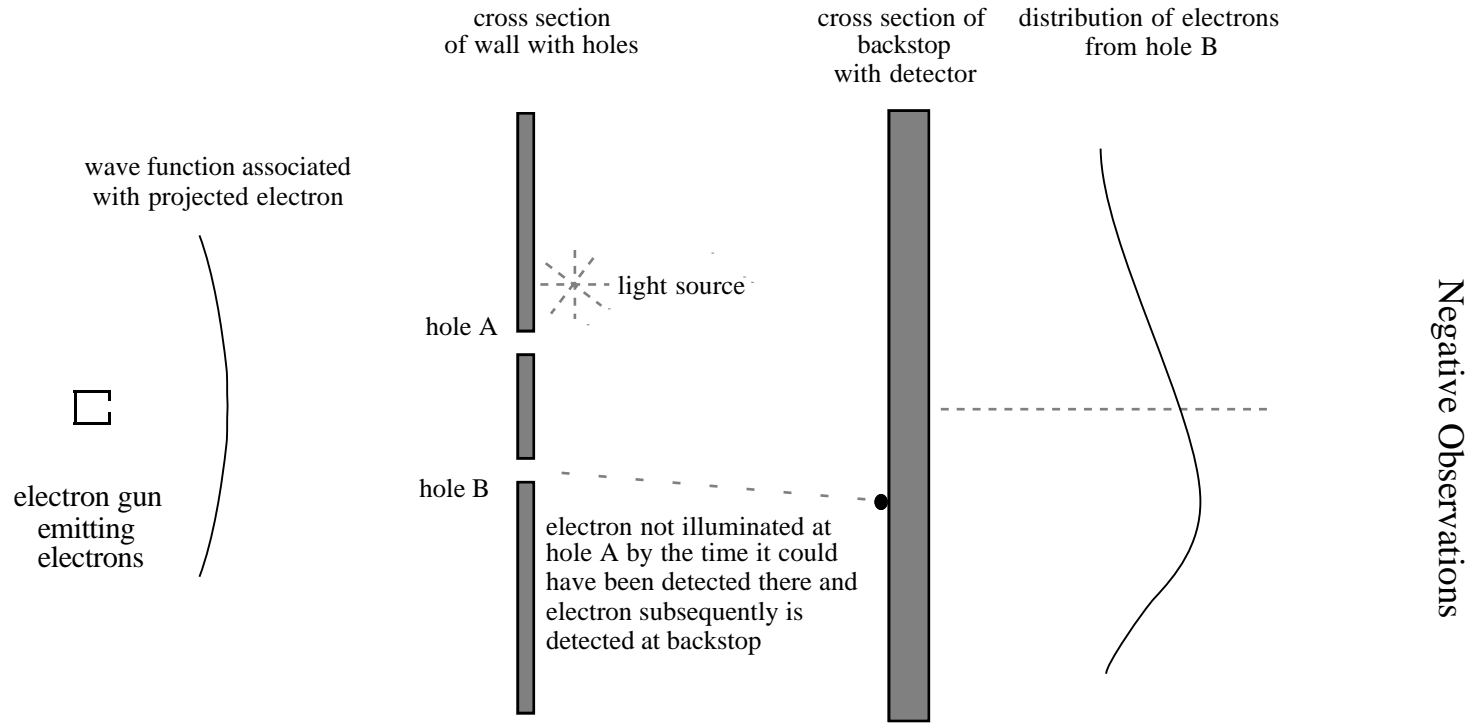

Figure 5

Two-hole gedankenexperiment with strong light source illuminating only one
hole in which the distribution of electrons from unilluminated hole is shown.

# Negative Observations

Epstein (1945) maintained that these kinds of different effects on the physical world in quantum mechanics that cannot be ascribed to physical causes are associated with "*mental certainty*" (p. 134) on the part of an observer as to which of the possible alternatives for a physical existent occurs. Indeed, the factor responsible for the change in the wave function for an electron headed for holes A and B, and which is not illuminated at hole A, is *knowledge* by the observer as to whether there is sufficient time for an electron to pass through the "illuminated" hole. To borrow a term used by Renninger (1960), when the time has elapsed in which the electron could be illuminated at hole A, and it is not illuminated, the observer makes a "negative" (p. 418) observation.

The common factor associated with the electron's passage through the wall in a manner resembling that found for classical-like particles in Gedanken-experiments 2 and 3 is the observing, thinking individual's knowledge as to whether an electron passed through a particular hole. The physical interaction between photons from the light source and electrons passing through either hole 1 or hole 2 is not a common factor. It should be remembered that some causative factor is implied by the very different electron distributions in Gedankenexperiments 1 and 2. It is reasonable to conclude that knowledge by the observer regarding the particular path of the electron through the wall is a factor in the change in the distribution of the electrons in Gedankenexperiment 1 to that found for electrons in Gedankenexperiments 2 and 3.

It might be argued that in Gedankenexperiment 3 a non-human recording instrument might record whether or not an electron passed through the illuminated hole in the time allowed, apparently obviating the need for a human observer. But, as has been shown, a non-human recording instrument is not necessary to obtain the results in Gedankenexperiment 3. And yet even if a non-human instrument is used, ultimately a person is involved to read the results who could still be responsible for the obtained results. Furthermore, one would still have to explain the destruction of the interference affecting the distribution of the electrons at the backstop without relying on a physical interaction between the electrons and some other physical existent. Without ultimately relying on a human observer, this would be difficult to accomplish when the non-human recording instrument presumably relies on physical interactions for its functioning.

It should also be emphasized that the change in the wave function for an electron passing through the unilluminated hole in Gedankenexperiment 3 provides the general case concerning what is necessary for the change in a wave function to occur in a measurement of the physical existent with which it is associated. It was shown clearly in the extension of Feynman et al.'s gedankenexperiments that the change in the wave function of an electron or other physical existent is not due fundamentally to a physical cause. Instead, the change in the wave function is linked to the knowledge attained by the observer of the circumstances affecting the physical existent measured.

There is one other point to be emphasized. The change in the wave function discussed in Gedankenexperiment 3 serves only to capture the role of knowledge in negative observation. *That is, one need not even present a discussion of the wave function to attain the result that knowledge is a factor in the change in the electron distribution in Gedankenexperiment 1 to the electron distribution in Gedankenexperiments 2 and 3. This result depends only on the analysis of experimental results concerning the electron distributions in these three gedankenexperiments.*

## The Schrödinger Cat
## Gedankenexperiment

The nature of the change in the wave function that generally occurs in a measurement will now be discussed in more detail in terms of a gedanken-experiment proposed in 1935 by Schrödinger. In his gedankenexperiment, Schrödinger focused on the immediate change in the wave function that occurs upon observation of a measuring apparatus that records the value of a quantum mechanical quantity.

A cat is penned up in a steel chamber, along with the following diabolical device (which must be secured against direct interference by the cat): in a Geiger counter there is a tiny bit of radioactive substance, so small, that perhaps in the course of one hour one of the atoms decays, but also, with equal probability, perhaps none; if it happens, the counter tube discharges and through a relay releases a hammer which shatters a small flask of hydrocyanic acid. If one has left this entire system to itself for an hour, one would say that the cat still lives if meanwhile no atom has decayed. The first atomic decay would have poisoned it. The $\Psi$-function of the entire system would express this by having in it the living and the dead cat (pardon the expression) mixed or smeared out in equal parts.

# Negative Observations

> It is typical of these cases [of which the foregoing example is one] that an indeterminancy originally restricted to the atomic domain becomes transformed into macroscopic indeterminancy, which can then be resolved by direct observation. (Schrödinger 1935/1983, p. 157)

How does the gedankenexperiment indicate that the nature of the wave function as a link between cognition and the physical world is warranted? It does so in terms of the features of the quantum mechanical wave function cited earlier, one being that there is no source of information concerning the physical world in quantum mechanics other than the probabilistic predictions that yield knowledge of the physical world, predictions that have been supported by empirical test. The second is that these probabilities in general change immediately throughout space upon observation of a quantity of the physical existent that is described by the wave function which is the basis for the probabilistic predictions. Importantly, the velocity limitation of the special theory *precludes a physical existent* from mediating this change in the wave function.

Note that Schrödinger does not specify how close the observer needs to be to the cat to resolve the indeterminancy. The observer can, in principle, be at any distance from the cat, even across the universe, and initiate this immediate change in the wave function, so long as the observer makes an observation regarding whether the cat is alive. Indeed, the observer does not even have to observe the cat directly but can rely on another observer who has observed the cat and who tells the former observer the result of his observation.

In a related vein, Schrödinger did not explicitly discuss the role and significance of the person as observer in the measurement process in quantum mechanics. Physicists often use the term "observation" ambiguously. Changing the latter part of Schrödinger's quote to indicate that the concern specifically is with a *person* making the observation does not lessen the statement's validity:

> It is typical of these cases [of which the foregoing example is one] that an indeterminancy originally restricted to the atomic domain becomes transformed into macroscopic indeterminancy, which can then be resolved by direct [human] observation.

Thus, in a circumstance where the observer is specified to be a person, the change in the wave function is tied explicitly to the perception by the human observer of the cat. This point is not limited to those circumstances where a human observer is explicitly specified. This point holds in the general case where a non-human macroscopic measuring instrument intervenes between a quantum mechanical entity and a human observer. It is a human observer who ultimately records the result of any observation. In the cat gedankenexperiment, for example, the cat acts as a macroscopic measuring instrument and comes to be characterized by the same probabilities as the microscopic physical phenomenon (i.e., the radioactive substance) until a human observer makes his own observation of the cat regarding its being alive or dead.

It should be remembered that the Schrödinger cat gedankenexperiment portrays the special case where a macroscopic measuring instrument is used to make a measurement. As has been shown, Gedankenexperiment 3 discussed above provides the general case concerning what is necessary for the change in a wave function to occur in a measurement of the physical existent with which it is associated. There it was also shown that the change in the wave function is linked to the knowledge attained by the observer of the circumstances affecting the physical existent measured and that the change in this wave function is not due fundamentally to a physical cause.

### Knowledge and the Measurement of the Spin Component of Electrons Along a Spatial Axis

It has been shown in gedankenexperiments using the two-hole interference scenario of Feynman, Leighton, and Sands that physical interaction is not necessary to effect the change in the wave function that generally occurs in measurement in quantum mechanics. Instead, the general case is that knowledge is linked to the change in the wave function. Another demonstration of this point follows. The models for gedankenexperiments employing electrons (spin one-half particles) presented now are found in Feynman, Leighton, and Sands's (1965) chapter on spin-one particles in their *Lectures on Physics*. Similar to the earlier gedankenexperiments, these gedanken-experiments also employ negative observation. But in contrast to the earlier gedankenexperiments, readily quantifiable results of the negative observations are developed. In addition, the significance of knowledge to the change of the wave function is emphasized because a concurrent physical interaction to the negative observation between the existent measured and the measuring instrument is shown to be incapable of effecting the change in the wave function.

# Negative Observations

*Basic Features of the Experimental Design*

Consider the case of a device like a Stern-Gerlach type apparatus (device A) which has an inhomogeneous magnetic field where the field direction and the direction of the gradient are the same, for example along the *z* axis (Figure 6). An electron can pass along one of two paths as it moves through the apparatus.[10] This is due to the quantization of the spin angular momentum of the electron, more specifically the quantization of the spin component along any spatial axis into two possible values.

Initially, let an electron be in a state such that the probabilities of its going through either of the paths are equal. Which of the two possible paths an electron has passed through depends on whether the electron's spin component along the axis of the inhomogeneous magnetic field of the device is either in, or against, the direction of the magnetic field and its gradient. Given the initial probabilities, one-half of the electrons exiting from device A will be observed to have spin up (i.e., in the direction of the magnetic field and gradient of device A), and one-half of the electrons exiting device A will be observed to have spin down (i.e., opposite to the direction of the magnetic field and gradient of device A). If, after an observation is made, the electron is now put through another Stern-Gerlach type device (device C), identical in construction to the first and oriented in the same direction, the electron will exit along the same path that it exited from in the first machine. In order to do this, the electron must first be brought back to its original direction of motion.

This is accomplished through the use of another Stern-Gerlach type device (device B), the spatial orientation of which is up-down and right-left reversed with respect to the first device. In device B, the magnetic field and the gradient are in the opposite direction along the same spatial axis to that found for device A. The placement of these two devices is shown in Figure 7, with devices A and B right next to each other.[11]

---

[10] An electron is a member of a class of particles known as fermions. The spin component of a fermion along any spatial axis has two possible values when it is measured: $+1/2$ ($h/2\pi$) (spin up along this axis) and $-1/2$ ($h/2\pi$) (spin down along this axis). The results of the gedankenexperiment hold for fermions in general.

[11] Note that no pathways are shown in Figure 7 for the electrons traveling through device AB. This is because quantum mechanics provides the correct description of the electrons, and it indicates that an electron does not travel over one or the other of the paths until an observation of the electron is made regarding which path it traveled. Instead, the wave function associated with an electron indicates that the probability is 1/2 that it will have spin up along the *z* axis and the probability is 1/2 that it will have spin down along the *z* axis when its spin

component along this axis is measured. In devices like AB in other gedankenexperiments where both paths are open, the lack of path lines will similarly indicate a lack of knowledge regarding which path electrons take in going through the device.

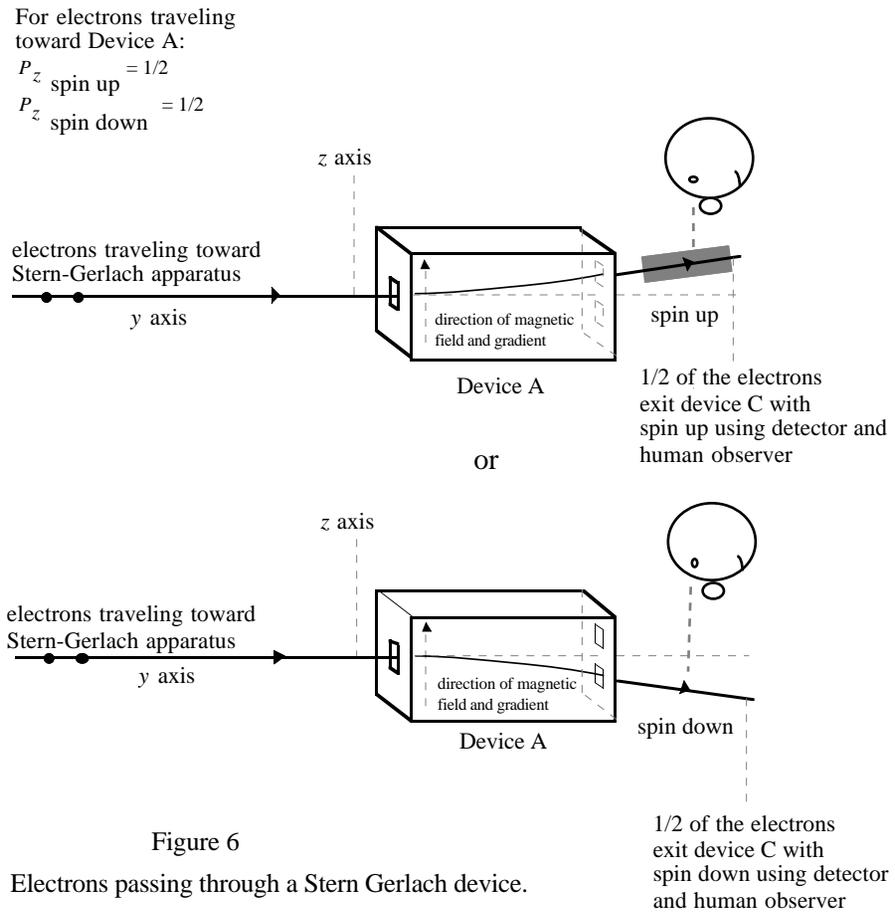

For electrons traveling toward Device A:

$P_z$ spin up $= 1/2$

$P_z$ spin down $= 1/2$

z axis

electrons traveling toward
Stern-Gerlach apparatus

y axis

direction of magnetic
field and gradient

Device A

spin up

1/2 of the electrons
exit device C with
spin up using detector and
human observer

or

z axis

electrons traveling toward
Stern-Gerlach apparatus

y axis

direction of magnetic
field and gradient

Device A

spin down

1/2 of the electrons
exit device C with
spin down using detector
and human observer

Figure 6

Electrons passing through a Stern Gerlach device.

For electrons traveling
toward Device A:
$P_{z \text{ spin up}} = 1/2$
$P_{z \text{ spin down}} = 1/2$

For electrons traveling
toward Device C:
$P_{z \text{ spin up}} = 1/2$
$P_{z \text{ spin down}} = 1/2$

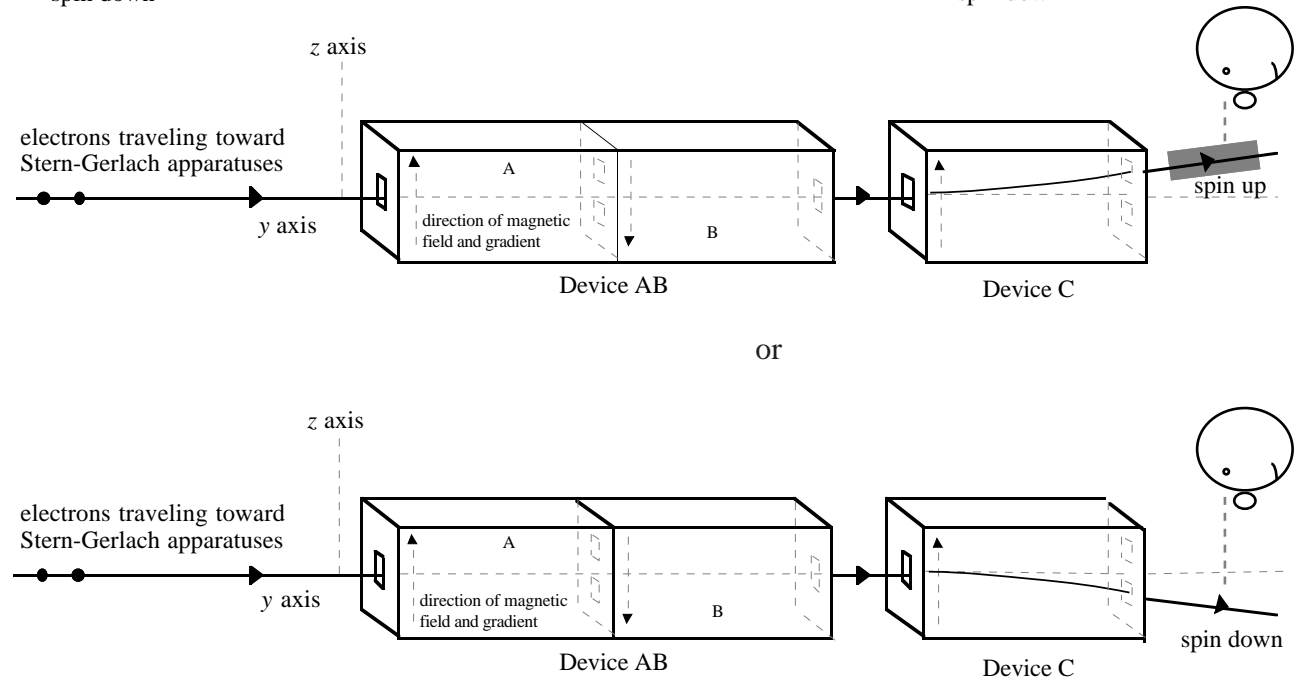

or

Figure 7

Electrons passing through a series of Stern-Gerlach
devices oriented along the same spatial axis $z$.

# Negative Observations

Consider the following gedankenexperiments that adhere to quantum mechanical principles and that are supported by empirical evidence. They show that it is an individual's knowledge of the physical world that is tied to the functioning of the physical world itself.

### Gedankenexperiment 4

Allow that device AB has a block inserted in it as portrayed in Figure 8. Then device AB allows only electrons with a spin up component along the $z$ axis to exit it. Electrons with a spin down component along this axis are blocked from exiting. Allow that $R$ electrons exit the device with a spin up component. Next to device AB a second device, DE, is placed that is identical in construction. D is the Stern-Gerlach-like device closest to B. The device DE is tilted around the $y$ axis relative to device AB. $aR$ electrons exit device DE with spin up (where $0 < a < 1$). (Spin up here is relative to the $z'$ axis and is in the direction of the magnetic field and gradient of device D.) Next to device DE is device C in the same spatial orientation as device A of AB and its magnetic field and gradient in the same direction along the $z$ axis as device A. A block is inserted into device C that precludes electrons with spin down from exiting it. $baR$ electrons exit device C with spin up (where $0 < b < 1$). (Spin up here is relative to the $z$ axis.) (Figure 9 displays the number of electrons exiting the various devices in this and succeeding gedankenexperiments.)

### Gedankenexperiment 5

The experimental arrangement is the same as that in Gedankenexperiment 4, except that no block is inserted in device DE (Figure 10). The numbers of electrons coming out of each device are as follows: (1) $R$ electrons exit device AB with spin up along the $z$ axis; (2) $R$ electrons exit device DE; and (3) $R$ electrons exit device C with spin up along the $z$ axis.

### Discussion of Gedankenexperiments 4 and 5

How can one account for the results of Gedankenexperiments 4 and 5? An observer finds that $R$ electrons exit device C in Gedankenexperiment 5, in accordance with the expectation that the spin components of the electrons along the $z$ axis remain unaffected by the passage of the electrons through device DE. It appears that device DE, which has no block, has no effect on the spin components along the $z$ axis of the electrons passing through it. $R$ electrons exit device A with spin up along the $z$ axis and $R$ electrons exit device C with

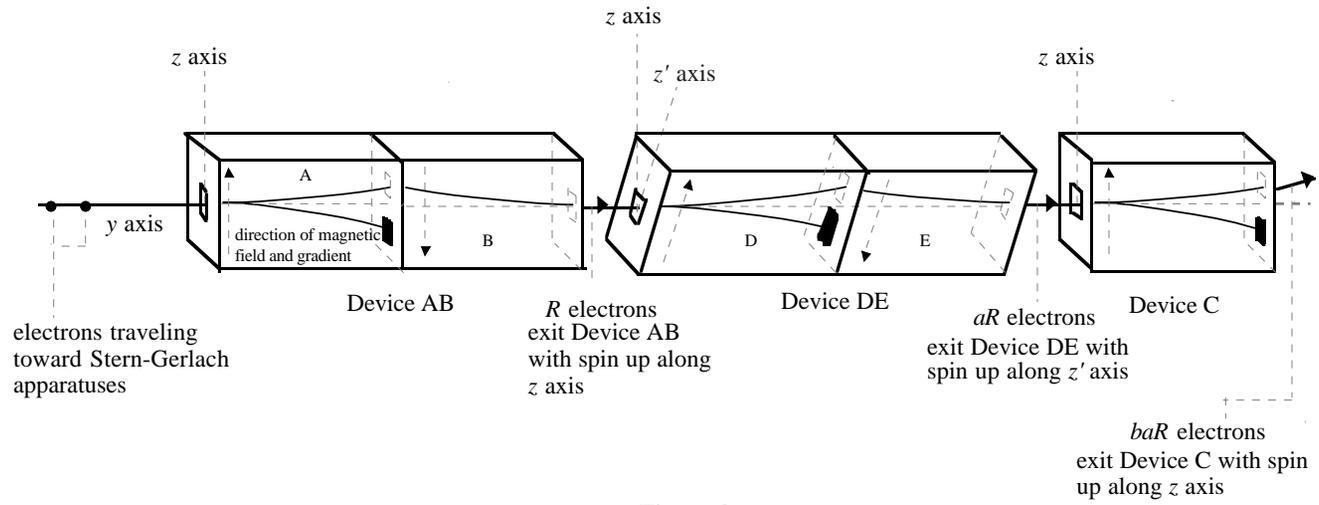

electrons traveling
toward Stern-Gerlach
apparatuses

Device AB

*R* electrons
exit Device AB
with spin up along
*z* axis

Device DE

*aR* electrons
exit Device DE with
spin up along *z′* axis

Device C

*baR* electrons
exit Device C with spin
up along *z* axis

Figure 8

A series of Stern-Gerlach devices where only electrons with
spin up along the *z* axis pass through device AB, only
electrons with spin up along the *z′* axis pass through device
DE, and only electrons with spin up along the *z* axis pass
through device C. (Gedankenexperiment 4)

| Gedanken-experiment | Device AB | | | Device DE | | | Device C | | |
|---|---|---|---|---|---|---|---|---|---|
| | Paths Open | Orientation of Magnetic Field | Number of Electrons Exiting AB | Paths Open | Orientation of Magnetic Field | Number of Electrons Exiting DE | Paths Open | Orientation of Magnetic Field | Number of Electrons Exiting C |
| 4 | spin up | $z$ axis | $R$ | spin up | $z'$ axis | $aR$ | spin up | $z$ axis | $baR$ |
| 5 | spin up | $z$ axis | $R$ | spin up and spin down | $z'$ axis | $R$ | spin up | $z$ axis | $R$ |
| 6 | spin up | $z$ axis | $R$ | spin down | $z'$ axis | $vR$ | spin down | $z$ axis | $uvR$ |
| 7 | spin up | $z$ axis | $R$ | spin up and spin down | $z'$ axis | $R$ | spin down | $z$ axis | $0$ |

Figure 9

Specifications and results for Gedankenexperiments 4 through 7.

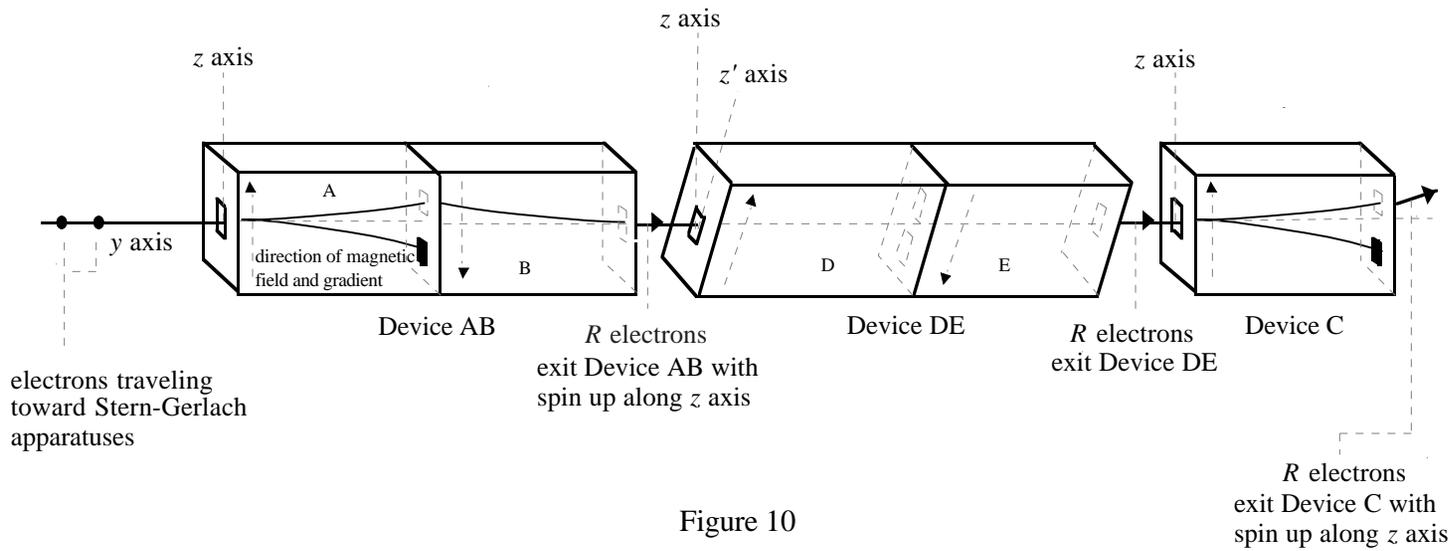

electrons traveling
toward Stern-Gerlach
apparatuses

Figure 10

A series of Stern-Gerlach devices where only electrons with spin up
along the *z* axis pass through device AB, paths in device DE for
electrons with spin up or spin down along the *z'* axis are both open, and
only electrons with spin up along the *z* axis pass through device C.
(Gedankenexperiment 5)

spin up along the *z* axis. All electrons pass through device DE. But Gedanken­experiment 4 does not provide a similar result. A similar result would be that *aR* electrons would exit device C in Gedankenexperiment 4, not *baR* electrons. That is, the spin components of the electrons along the *z* axis would essentially remain unaffected by the passage of the electrons through device DE in Gedankenexperiment 4, just as device DE in Gedankenexperiment 5 does not appear to affect the spin components of electrons along the *z* axis. How is it that *baR* electrons exit from device C in Gedankenexperiment 4 instead of *aR* electrons? It is reasonable to conclude that something unusual is happening to the electrons in their passage through device DE in Gedankenexperiment 4, particularly in view of the results of Gedankenexperiment 5. Somehow the spin components of the electrons along the *z* axis are affected by their passage through device DE in Gedankenexperiment 4 while device DE in Gedanken­experiment 5 does not affect the spin components of electrons along the *z* axis.

A comparison of Gedankenexperiments 4 and 5 indicates that the only physical feature of the measuring apparatus that can possibly be responsible for the change in the component of the spin angular momentum along the *z* axis of the electron is the block that is inserted in device DE in Gedankenexperiment 4. Other than this one difference, the measuring apparatuses in Gedanken­experiments 4 and 5 are identical.

### The Block in Device DE

The experimental consequences resulting from the presence or absence of the block in device DE in Gedankenexperiments 4 and 5 concern whether one or both paths are open in device DE. *Significantly, it is electrons traveling along the unblocked path in Gedankenexperiment 1 that exhibit the unusual behavior regarding the frequency of electrons exiting device C*. Thus, the nature of the effect of the influence of the block on the electrons is indeed unusual from a conventional standpoint, a standpoint that would expect the change in spin components along the *z* axis of the electrons that travel along the unblocked path to somehow be changed by a physical interaction with the block. This physical interaction, though, is not possible. The scenario involving a block is thus in essence a negative observation. A negative observation occurs where an observation is made by deducing that a particular physical event must have occurred because another physical event did not occur with subsequent consequences for the functioning of the physical world stemming from the change in knowledge. Physical interaction as the basis for the consequences in the physical world is ruled out. Remember that the spin components of the electrons along the $z$ axis traveling through device DE are affected by the change in knowledge, as evidenced by *baR* electrons exiting device C in Gedankenexperiment 4 instead of *aR* electrons. As previously noted, empirical work on electron shelving that supports the existence of negative observation has been conducted by Nagourney, Sandberg, and Dehmelt (1986), Bergquist, Hulet, Itano, and Wineland (1986), and by Sauter, Neuhauser, Blatt, and Toschek (1986).

*A Variation of the Gedankenexperiments*

Two other gedankenexperiments similar to Gedankenexperiments 4 and 5 will provide an even more remarkable demonstration that an individual's knowledge of the physical world is tied to the functioning of the physical world itself.

Gedankenexperiment 6

The experimental arrangement is the same as that in Gedanken-experiment 4, except that the blocks are inserted in devices DE and C such that spin up electrons along $z'$ and $z$, respectively, cannot exit these devices and spin down electrons are allowed to proceed unimpeded (Figure 11). The numbers of electrons coming out of each device are as follows: (1) $R$ electrons exit device AB with spin up along the $z$ axis; (2) $vR$ electrons exit device DE with spin down along the $z'$ axis; and (3) $uvR$ electrons exit device C with spin down along the $z$ axis.

Gedankenexperiment 7

The experimental arrangement is the same as that in Gedanken-experiment 6, except that device DE has both paths open (Figure 12). The numbers of electrons coming out of each device are as follows: (1) $R$ electrons exit device AB with spin up along the $z$ axis; (2) $R$ electrons exit device DE; and (3) 0 electrons exit device C with spin down along the $z$ axis.

Discussion of Gedankenexperiments 6 and 7

The result in Gedankenexperiment 6 is remarkable. How is it that electrons with spin down along the axis of the magnetic field of the measuring device A, oriented in a particular direction along $z$, are found exiting device C, in which the axis of its magnetic field and its gradient are also oriented in the same direction along $z$? No electron with spin down along the $z$ axis exits device AB. This result is particularly unusual when in Gedankenexperiment 7,

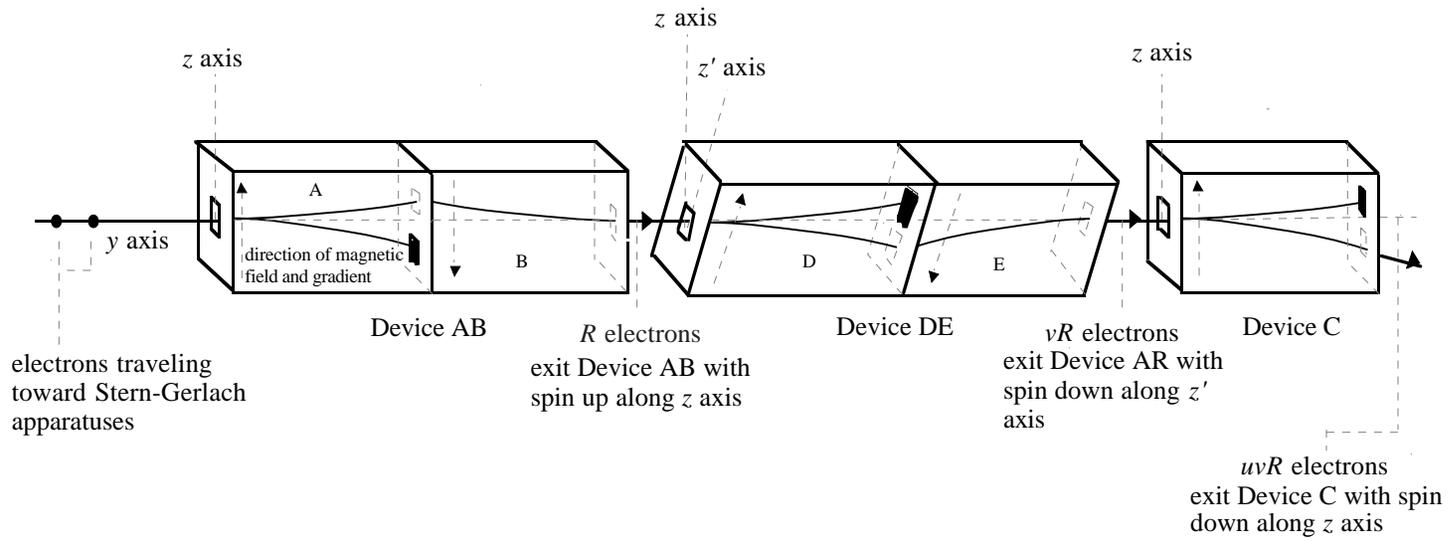

z axis

z axis
z′ axis

z axis

y axis

A

direction of magnetic
field and gradient

B

Device AB

D

E

Device DE

Device C

electrons traveling
toward Stern-Gerlach
apparatuses

R electrons
exit Device AB with
spin up along z axis

vR electrons
exit Device AR with
spin down along z′
axis

uvR electrons
exit Device C with spin
down along z axis

Figure 11

A series of Stern-Gerlach devices where only electrons with spin up along
the z axis pass through device AB, only electrons with spin down along the
z′ axis pass through device DE, and only electrons with spin down along the
z axis pass through device C.  (Gedankenexperiment 6)

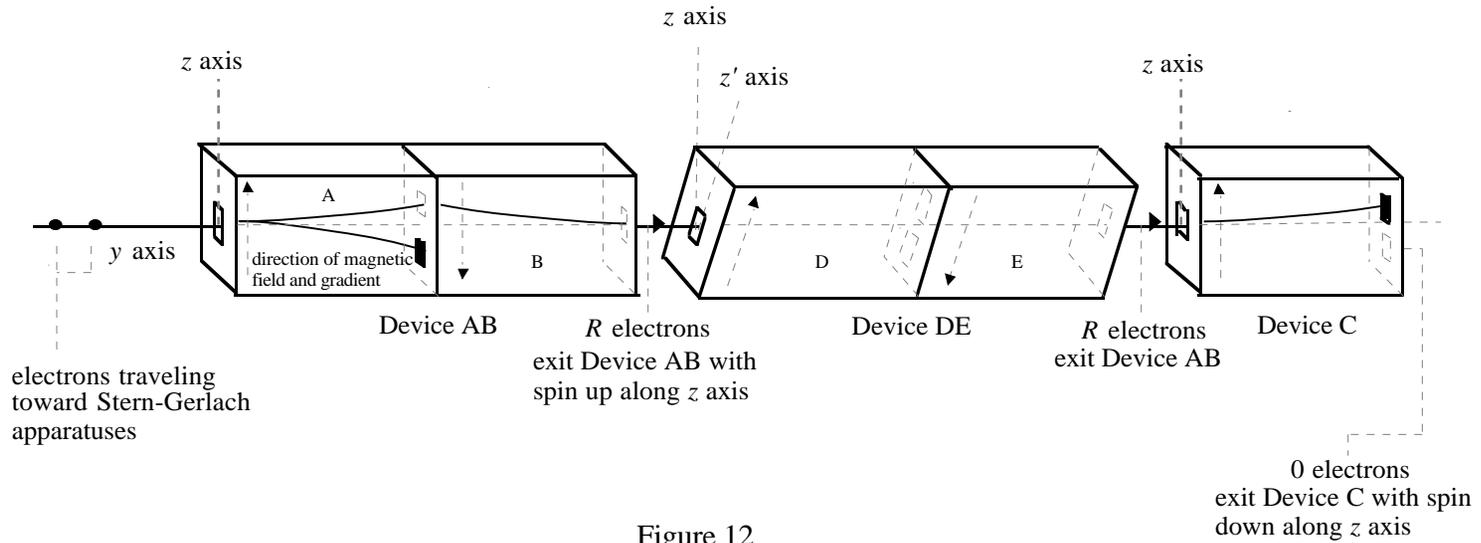

Figure 12

A series of Stern-Gerlach devices where only electrons with spin up along the
$z$ axis pass through device AB, paths in device DE for electrons with spin up
or spin down along the $z'$ axis are open, and only electrons with spin down
along the $z$ axis pass through device C.  (Gedankenexperiment 7)

using the same device DE, modified only by the removal of the block that prevents electrons with a spin up component along $z'$ (the axis of the magnetic field in DE) to pass, there are no electrons that exit device C with spin down along the axis of its magnetic field, which has the same spatial orientation as the magnetic field of A along the $z$ axis.

In Gedankenexperiment 7, it appears as if the spin components of the electrons along the $z$ axis were not affected by their passage through device DE, which has both paths open and which thus allowed all electrons to pass through. As reflected in the behavior of the electrons that pass through device C, the spin components of the electrons along the $z$ axis in Gedankenexperiment 6 are affected by device DE, specifically by the insertion of the block in this device that prevents electrons with spin up components along the $z'$ axis from exiting device DE. Again, no electrons with spin down along this axis were found to exit device AB. *The electrons traveling along the unblocked path in device DE in Gedankenexperiment 6 exhibit this unusual behavior regarding the frequency of electrons exiting device C.* No physical interaction between the block in device DE and any electron traveling along the unblocked path is responsible for the frequency of electrons exiting device C. In Gedanken-experiment 6, a negative observation at device DE has resulted in electrons exiting device C with spin down along the $z$ axis whereas in the absence of a negative observation, in Gedankenexperiment 7, no electrons exit device C with spin down along the $z$ axis.

*Interference*

The difference in the observer's knowledge of the spin components of electrons along an axis, and the difference in the spin components of the electrons themselves, in the pairs of gedankenexperiments that have been presented (i.e., Gedankenexperiments 4 and 5, and 6 and 7) reflect the presence or absence of interference in the wave functions associated with each of the electrons. For example, in terms of the formalism, in Gedankenexperiment 4 the probability amplitude $a_1$ for an electron exiting device AB with spin up (*AB+*) and exiting device C with spin up (*C+*) is given by

$$a_1 = <C+|DE+> <DE+|AB+> . \qquad (21)$$

The probability of these events is derived by taking the absolute square of this probability amplitude, $|a_1|^2$. In contrast, in Gedankenexperiment 5, the probability amplitude $d$ for an electron exiting device AB with spin up (*AB+*) and exiting device C with spin up (*C+*) is given by

$$d = <C+|DE-> <DE-|AB+> + <C+|DE+> <DE+|AB+> \quad . \tag{22}$$

When the absolute square of the probability amplitude $d$ is calculated to yield the probability that an electron exiting device AB with spin up will exit device C with spin up, it is evident that there will be two terms representing interference. These terms are

$$(<C+|DE-> <DE-|AB+>)^* \ (<C+|DE+> <DE+|AB+>) \tag{23a}$$

and

$$(<C+|DE+> <DE+|AB+>)^* \ (<C+|DE-> <DE-|AB+>) \quad . \tag{23b}$$

It is these terms that distinguish $|d|^2$, where there is interference, from $\Sigma |a_i|^2$ where one knows which path the electron took through device DE and there is no interference

$$\Sigma |a_i|^2 = |a_1|^2 + |a_2|^2 \tag{24}$$

or

$$\Sigma |a_i|^2 = |<C+|DE+> <DE+|AB+>|^2 + |<C+|DE-> <DE-|AB+>|^2 \quad . \tag{25}$$

It is important to emphasize that it is not the presence or absence of the block in device DE that interacts with electrons that is responsible for the presence or absence of interference in Gedankenexperiments 4 and 6. It is the act of knowing the value of the spin component of the electron along the $z'$ axis that is responsible. The block in device DE in Gedankenexperiment 4 and the block in device DE in Gedankenexperiment 6 serve as bases for negative observations.

### Another Indication of the Importance of Knowledge in Measurement in Quantum Mechanics

There is one more feature of the gedankenexperiments discussed in this paper that supports the theses that: 1) the macroscopic nature of a physical apparatus used for a measuring instrument is not central to making a measurement in quantum mechanics; 2) knowledge is central to making such measurements; and 3) the role of the block in device DE in Gedanken-experiments 4 and 6 is to provide information. Gedankenexperiments 4 through 7 demonstrate the interesting point that the magnetic field of device DE itself is not sufficient to induce the change in the wave function that device DE which has a block along one path does for electrons traveling along the unblocked path and with which the block does not physically interact. Unless there is some way in the physical set up of device DE to determine the spin component of the electron along an axis $z'$ (as is done in Gedankenexperiments 4 and 6 by the block in device DE), there is no change in the wave function of the electron concerning its spin components.

In Gedankenexperiment 5 where there is no possibility in the physical set up that is device DE to know the spin component of the electron in device DE (because a block is not inserted along either the "spin up" or the "spin down" path), device DE does not affect the spin components along the $z$ axis of the electrons as they travel through. That is, the number of electrons exiting devices C and AB are exactly the same. Also, in gedankenexperiment 7, no electrons with spin down along the $z$ axis exit device C and only electrons with spin up along the $z$ axis exit device AB. Without a block in device DE in Gedankenexperiment 7, there is no change in the wave function of an electron as regards its spin components. This is equivalent to saying that there has been no measurement of the spin component along the $z'$ axis of the electron. To quote Feynman et al. (1965) regarding their filtering experiments with spin-one particles similar in principle to Gedankenexperiments 4 and 6:

> The past information [concerning spin along the $z$ axis after exiting the first device] is not lost by the *separation* into...beams [in the second device], but by the *blocking masks* that are put in [the second device] (p. 5-9).

In conclusion, if an interaction between a macroscopic physical apparatus and the existent to be measured were responsible for a change in the wave function of the physical existent measured, why, if a magnetic field by itself is unable to effect this change in the wave function for electrons, is the insertion of a block able to effect this change for electrons *traveling through the unblocked path*? When device DE does not contain a block along one of the paths, electrons traveling along what is the unblocked path in Gedankenexperiment 4 or the unblocked path in Gedankenexperiment 6 do not undergo any change in their wave function. The role of the block in Gedankenexperiments 4 and 6 is to provide information to a human observer concerning electrons traveling along the unblocked path. With regard to these electrons, the role of the block in the measurement of their spin components along the $z'$ axis does not depend on a physical interaction between them and the block.

# Negative Observations

## The Time of a Measurement

The question is often asked concerning quantum mechanics how can an observer finding out about a measurement that has presumably been made some time earlier be linked to the measurement itself? In terms of Gedanken-experiment 4, for example, if a human observer finds out about the electrons passing through the devices AB, DE, and C only after the electrons exit device C, how can this observer be considered responsible in some way for a measurement that was presumably made at device DE because of the inclusion of the block in that device? That is, a negative observation seems to be made only after the electrons exit device C, even though the block in device DE made the information available earlier (i.e., as soon as the time elapsed in which an electron passing through device DE could reach the block at the end of D).

The analysis underlying the question presumes that some form of physical interaction occurring within a temporal framework provides the basis for measurement in quantum mechanics even though it clearly does not. In Gedankenexperiment 4, this presumed physical interaction does not occur in device DE. Measurement in quantum mechanics is *fundamentally* concerned with the development of knowledge. The course of physical interactions over time is not the central factor in the development of this knowledge. It is knowledge that is primary and within this knowledge, the functioning of the physical world, including the course of physical interactions over time, occurs.

As has been discussed, there are other indications for this view concerning the importance of knowledge in quantum mechanics. Knowledge of the physical world is developed using wave functions, and wave functions provide only probabilistic knowledge. The quantum mechanical wave function associated with a physical existent generally changes immediately throughout space upon measurement of the physical existent. This change in the wave function is not limited by the velocity limitation of the special theory of relativity for physical existents, the velocity of light in vacuum. There is the complex number nature of the wave function from which information concerning the physical world is derived.

## The Effect of Measurement on the Past

One other point provides support for the central significance of knowledge in measurement in quantum mechanics. In Gedankenexperiments 4 and 6, the presence of the block, or more accurately the knowledge that results from the presence of the block, at the exit of device D affects the electrons traveling along the unblocked path in device D from their *entry* into device D for two reasons:

1. If the block is removed prior to the end of the time over which an electron could traverse device D along the blocked path, interference would not be destroyed and the number of electrons exiting device C in Gedankenexperiment 4 (i.e., with spin up along the *z* axis), for example, is the same as the number of electrons exiting device AB.

2. With the block in place and the time elapsed over which an electron could have reached the block in device D, the interference that was supposed to characterize the electron in its passage through device D did not occur as the electron could have traveled along only the unblocked path. If a detector had been set up along any part of the path in device D containing the block prior to the electron's having reached the end of device D where the block is situated, the electron would not have been detected along the path containing the block.

A negative observation that the block allows for by providing information to an observer is thus seen to affect one's knowledge of the past as well as the past itself, in the present case indicating that the electron has traveled down a particular path in device D as opposed to being characterized by a wave function demonstrating interference and not having traveled one path exclusively.

## References

Bergquist, J. C., Hulet, R. G., Itano, W. M., and Wineland, D. J. (1986). Observation of quantum jumps in a single atom. *Physical Review Letters*, *57*, 1699-1702.

Bohr, N. (1935). Can quantum-mechanical description of nature be considered complete? *Physical Review*, *49*, 1804-1807.

Cook, R. J. (1990). Quantum jumps. In E. Wolf (Ed.), *Progress in Optics* (Vol. 28) (pp. 361-416). Amsterdam: North-Holland.

Dicke, R. H., and Wittke, J. P. (1960). *Introduction to quantum mechanics*. Reading, Massachusetts: Addison-Wesley.

Eisberg, R., and Resnick, R. (1985). *Quantum physics of atoms, molecules, solids, nuclei and particles* (2nd ed.). New York: Wiley. (Original work published 1974)

Epstein, P. (1945). The reality problem in quantum mechanics. *American Journal of Physics*, *13*, 127-136.

# Negative Observations

Feynman, P. R., Leighton, R. B., and Sands, M. (1965). *The Feynman lectures on physics: Quantum mechanics* (Vol. 3). Reading, Massachusetts: Addison-Wesley.

Gasiorowicz, S. (1974). *Quantum physics*. New York: John Wiley.

Goswami, A. (1992). *Quantum mechanics*. Dubuque, Iowa: Wm. C. Brown.

Liboff, R. (1993). *Introductory quantum mechanics* (2nd ed.). Reading, Massachusetts: Addison-Wesley.

Mermin, N. D. (1985, April). Is the moon there when nobody looks? Reality and the quantum theory. *Physics Today*, 38-47.

Merzbacher, E. (1970). *Quantum mechanics* (2nd. ed.). New York: John Wiley. (Original work published 1961)

Messiah, A. (1965). *Quantum mechanics* (2nd ed.) (Vol. 1) (G. Tremmer, Trans.). Amsterdam: North-Holland. (Original work published 1962)

Nagourney, W., Sandberg, J., and Dehmelt, H. (1986). Shelved optical electron amplifier: observation of quantum jumps. *Physical Review Letters*, *56*, 2797-2799.

Renninger, M. (1960). Messungen ohne Störung des Meßobjekts [Observations without disturbing the object]. *Zeitschrift für Physik*, *158*, 417-421.

Sauter, T., Neuhauser, W., Blatt, R. and Toschek, P. E. (1986). Observation of quantum jumps. *Physical Review Letters*, *57*, 1696-1698.

Schrödinger, E. (1983). The present situation in quantum mechanics. In J. A. Wheeler and W. H. Zurek, *Quantum theory and measurement* (pp. 152-167) (J. Trimmer, Trans.). Princeton, New Jersey: Princeton University Press. (Original work published 1935)

Snyder, D. M. (1990). On the relation between psychology and physics. *The Journal of Mind and Behavior*, *11*, 1-17.

Snyder, D. M. (1992). Quantum mechanics and the involvement of mind in the physical world: A response to Garrison. *The Journal of Mind and Behavior*, *13*, 247-257.

Snyder, D. M. (1996a). Cognition and the physical world in quantum mechanics. Paper presented at the annual convention of the Western Psychological Association, San Jose, California.

Snyder, D. M. (1996b). On the nature of the change in the wave function in a measurement in quantum mechanics. Los Alamos National Laboratory E-Print Physics Archive (WWW address: http://xxx.lanl.gov/abs/quant-ph/9601006).

Wigner, E. (1983). Remarks on the mind-body question. In J. A. Wheeler and W. H. Zurek, Quantum *theory and measurement* (pp. 168-181). Princeton, New Jersey: Princeton University Press. (Original work published 1961)